\documentclass[twocolumn]{aastex63}
\usepackage{verbatim}
\usepackage{comment} 
\def\be{\begin{equation}}
\def\ee{\end{equation}}

\newcommand\quotes[1]{``{#1}"}
\def\gsim{\lower.5ex\hbox{\gtsima}} 
\def\lsim{\lower.5ex\hbox{\ltsima}} 
\def\gtsima{$\; \buildrel > \over \sim \;$} 
\def\ltsima{$\; \buildrel < \over \sim \;$} \def\gsim{\lower.5ex\hbox{\gtsima}} 
\def\lsim{\lower.5ex\hbox{\ltsima}} 
\def\simgt{\lower.5ex\hbox{\gtsima}} 
\def\simlt{\lower.5ex\hbox{\ltsima}}

\def\msun{{\rm M}_{\odot}}

\def\zsun{{\rm Z}_{\odot}}

\newcommand{\angstrom}{\mbox{\normalfont\AA}}

\def\S*{$\Sigma_{\rm SFR}$}

\def\OIII{\hbox{[O~$\scriptstyle\rm III $]~}}
\def\ks{\kappa_{\rm s}}

\makeatletter
\def\@hex@@Hex#1%
 {\if a#1A\else \if b#1B\else \if c#1C\else \if d#1D\else
  \if e#1E\else \if f#1F\else #1\fi\fi\fi\fi\fi\fi \@hex@Hex}
\makeatother
\definecolor{afcolor}{HTML}{b3443c}

\revised{\today}
\shorttitle{Super-early blue galaxies}
\shortauthors{Ziparo, Ferrara, Sommovigo, Kohandel}
\graphicspath{{./}{figures/}}
\begin{document}

\title{Blue monsters. Why are JWST super-early, massive galaxies so blue?}

\correspondingauthor{Francesco Ziparo}
\email{francesco.ziparo@sns.it}

\author[0000-0001-6316-1707]{Francesco Ziparo}
\affil{Scuola Normale Superiore,  Piazza dei Cavalieri 7, 50126 Pisa, Italy}

\author[0000-0002-9400-7312]{Andrea Ferrara}
\affil{Scuola Normale Superiore,  Piazza dei Cavalieri 7, 50126 Pisa, Italy}

\author[0000-0002-2906-2200]{Laura Sommovigo}
\affil{Scuola Normale Superiore,  Piazza dei Cavalieri 7, 50126 Pisa, Italy}

\author[0000-0003-1041-7865]{Mahsa Kohandel}
\affil{Scuola Normale Superiore,  Piazza dei Cavalieri 7, 50126 Pisa, Italy}

\begin{abstract}
The recent JWST tentative discovery of a population of super-early (redshift $z> 10$), relatively massive (stellar mass $M_* = 10^{8-9} \msun$) and evolved (metallicity $Z \approx 0.1 Z_\odot$) galaxies, which nevertheless show blue ($\beta \simeq -2.6$) spectra, and very small dust attenuation ($A_{\rm V} \simlt 0.02$), challenges our interpretation of these systems. To solve the puzzle we propose two solutions in which dust is either (a) ejected by radiation pressure, or (b) segregated with respect to UV-emitting regions. We clarify the conditions for which the two scenarios apply, and show that they can be discriminated by ALMA observations, such as the recent non-detection of the 88$\mu$m dust continuum in GHZ2 ($z\simeq 12$) favouring dust ejection.   
\end{abstract}


\section{Introduction} \label{sec:intro}

%\begin{itemize}

The \textit{James Webb Space Telescope} (JWST) has already unveiled tens of bright ($M_{\rm UV}\sim -21$), massive and blue galaxy candidates at unprecedentedly high redshift \citep[$z>10$, see][]{Santini22,Adams22, Donnan22, Naidu22a, Finkelstein22, Castellano22, Atek22, Whitler22, Harikane22, Furtak22}. Even more strikingly, \citet{Windhorst22, Yan22} reported the detection of $\sim 20$ sources potentially located at $z>11$, some of which show extremely bright $M_{\rm UV}\simlt-23$. The first spectroscopically confirmed galaxies (4) around $z\sim 11.5$ have been reported by \citet{Robertson22}. For these sources, the photometric and spectroscopic redshifts only differ by $\leq 3\%$, which is very promising for the numerous other JWST-detected candidates with photometric $z \geq 10$.
The unexpected abundance of such super-early \quotes{blue monsters} is challenging the predicting power of standard theoretical models \citep{Dayal14, Dayal22, Behroozi19, Mason22, Boylan-Kolchin22,Lovell22}. 

Thanks to JWST new data, it has become possible to extend UV Luminosity Function (LF) studies up to  $z\simeq14$. \cite{Naidu22a, Donnan22} find a minimal evolution of the bright-end of the UV LF between $z=7$ and $z=14$. These results are in contrast with the extrapolation of $z\simeq 7$ LFs \citep{Bouwens16}, which would predict a sharp drop in the number density of sources at $M_{\rm UV}\simlt -21$.
    
Such discrepancy can be reconciled in at least two alternative ways. The first solution invokes a much higher star formation efficiency ($\epsilon_*\simeq 0.1-0.3$, \citealt{Mason22, Inayoshi22}) with respect to lower-$z$ systems. This would imply that JWST is probing relatively small, and thus abundant, halos ($M_h\simeq 10^{9-10}\ \mathrm{M_\sun}$), featuring surprisingly large stellar masses. 

A possibly simpler hypothesis is that galaxies at $z>10$ are essentially unattenuated by dust \citep{Ferrara22a}. Indeed, a drastic drop in the dust  optical depth 
results in much brighter galaxies, virtually compensating for the lower abundance of their host
halos. 
    
As already hinted, most of the observed sources at $z\simgt 10$ \citep{Adams22, Atek22, Castellano22, Finkelstein22, Naidu22a} are characterised by blue UV slopes, {in the range $-2.0 \simlt \beta \simlt -2.6$}. At face value, this evidence supports the hypothesis that early galaxies suffer very little obscuration.
Indeed, from the SED-fitting analysis of 15 gravitationally lensed $z=10-16$ galaxies behind the galaxy cluster SMACS J0723.3-7327, \citet[][]{Furtak22} found an upper limit to dust attenuation of $A_{\rm V} < 0.02$.

On the other hand, {most} JWST-detected galaxies appear to be relatively massive ($M_* = 10^{8-9} \msun$), and metal-enriched as indicated by their (a) blue ($\beta \simgt -2.6$), but not ultra blue\footnote{A moderately red $\beta$ slope can be produced by either dust attenuation or nebular emission even starting from a metal-free (Pop III), very blue stellar population, see \citet[][]{Bouwens10}.} ($\beta \simeq -3$) UV spectral slopes \citep{Topping22a, Cullen22,Furtak22,Naidu22a,Finkelstein22}; (b) low, but not primordial metallicities, $Z \simeq 0.1\, \zsun$ \citep{Furtak22}.

These two sets of evidences seem to point in opposite directions. While from the high stellar mass/moderate metallicity one would expect a consistently high dust content, the extremely low dust optical depth (e.g. $A_{\rm V}\approx 0.01$) implies either very low dust content or attenuation.

Here we propose two alternative solutions to reconcile the above tension: (a) dust ejection by radiation pressure, or (b) dust spatial segregation with respect to the UV emitting regions. These two scenarios carry very different implications for the FIR continuum emission that can be tested with ALMA observations.

\section{Preliminary considerations}\label{sec:prel}
The JWST discovery of a large number of super-early, massive galaxies at $z>10$ represents a theoretical challenge for most {structure formation} models. A possible solution \citep{Ferrara22a}
involves an extremely low attenuation of UV light by dust in these systems, that almost exactly compensates for the increasing shortage of their host halos, and keeps their observed abundance constant with redshift. 

Importantly, this hypothesis is supported by the surprisingly little attenuation, and frequent blue colors of these objects \citep[][]{Furtak22}. For example, Maisie's galaxy at $z\approx 12$ \citep{Finkelstein22} has $A_{\rm V}=0.06$; SMACS-z12a ($z=12.03$) has extremely steep UV spectral slope, $\beta=-2.71$ \citep{Atek22}.    {GHZ2 ($z=12.35$) shows a very blue slope $\beta = -3.00 \pm 0.12$ \citep{Castellano22, Santini22} based on the updated photometry based on the latest NIRCam calibration files; this value is in contrast with the one initially inferred by \cite{Bakx22} for the same source (there referred to as GL-z13 as in \citealt{Naidu22a}) $\beta = -2.4 \pm 0.13$. In the rest of the work we show the predictions obtained for both values of $\beta$ inferred for this source.}
Even more surprisingly, the galaxy 10234 at $z=11.49$ \citep{Adams22} is the bluest with a $\beta \approx -3.35$ that is even reminiscent of peculiar stellar populations\footnote{We warn that at least some of these extremely blue values might be spuriously produced by an observational bias, pushing measurements towards artificially blue $\beta$ values for faint sources near the detection threshold \citep{Cullen22}.}. 

\begin{figure}
\centering\includegraphics[width = 0.46 \textwidth]{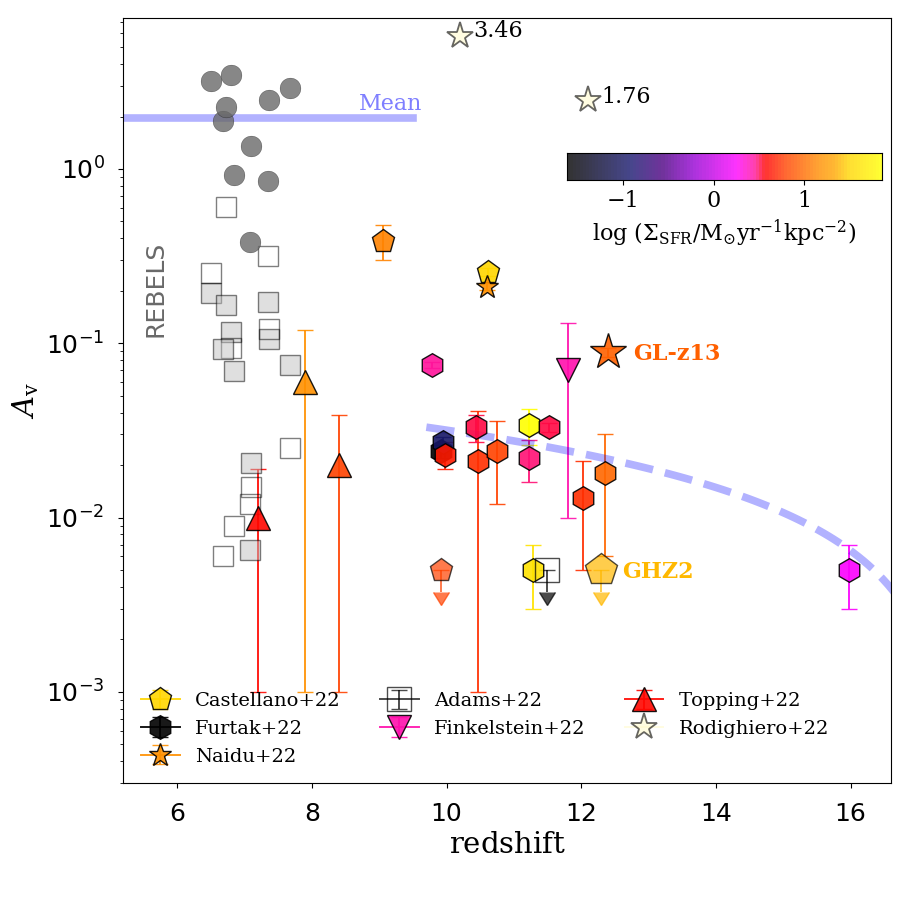}
\caption{Dust attenuation, $A_{\rm V}$, as a function of redshift for JWST-detected galaxies \citep[$z>9.5$,][]{Adams22, Castellano22, Finkelstein22, Furtak22,Naidu22a, Robertson22, Topping22} compared to ALMA and HST-detected galaxies from the REBELS survey \citep[$z\sim 7$, see:][]{Bouwens22}. 
JWST-detected galaxies are color-coded according to their SFR surface density $\Sigma_{\rm SFR}$ (when available). We also show the two $z>10$ JWST-detected sources from \cite{Rodighiero22}, which feature surprisingly large $A_{\rm V}$ (their $\log \Sigma_{\rm SFR}$ values are also shown). The blue dashed curve shows the best-fitting $A_{\rm V}-z$ relation at $z>9.5$; the solid one represents the mean $A_{\rm V}$ for ALMA-detected REBELS galaxies inferred by consistently modelling both their UV and FIR emission \citep{Ferrara22}. For comparison, also shown are the $A_{\rm V}$ values for the same galaxies derived from UV-to-NIR SED fitting only, and assuming either a constant (empty squares, Stefanon et al., in prep) or non-parametric \citep[grey squares,][]{Topping22a} star formation history.   
}
\label{Fig:01}
\end{figure}

Such  decreasing trend of $A_{\rm V}$ with redshift is clearly seen in Fig. \ref{Fig:01}, where we have also collected, for comparison, data at $z \approx 7$ from the ALMA Large Program REBELS \citep{Bouwens22}. 
%Spiegazione derivazione A_{\rm V}
{Where available, we show $A_{\rm V}$ obtained from SED fitting \citep{Furtak22,Topping22,Finkelstein22}. For the other sources, we derive the dust attenuation from the observed UV slope using the standard relation $A_{\rm V}=0.41\ \tau_{1500}$ \citep[assuming MW dust,][]{Ferrara22}, where $\tau_{1500}\simeq (\beta - \beta_{\rm int})$ in the optically-thin limit. We assume an intrinsic UV slope $\beta_{\rm int}=-2.616$ \citep{Reddy18}; in the few cases where $\beta<\beta_{\rm int}$ we assign an upper limit for $A_{\rm V}\sim 0.005$ (which corresponds to the lowest value measured in $z>10$ JWST-detected galaxies, see \citealt{Furtak22}).} 

Although a decreasing dust content could be expected at early epochs as a result of the lower cosmic stellar mass\footnote{Stars, and in particular SNe \citep{Todini01} are the main dust factories for cosmic ages $\simlt 1$ Gyr.} density, $\Omega_*(z=8) \approx 10^{-3} \Omega_*(z=0)$ \citep{Song16}, the result remains puzzling.

A simple calculation serves to illustrate the problem. To fix ideas, consider the galaxies GHZ2/GLz13
%\footnote{This source has been assigned these two names by \citet[][]{Castellano22} and \citet[][]{Naidu22a}, respectively. In the following, for brevity, we will denote it with GHZ2.} 
and GLz11 \citep{Naidu22a, Castellano22}, whose mean stellar mass is $M_* = 10^{9.2} \msun$. If $\nu_{\rm SN} = (52.89 \msun)^{-1}$ supernovae (SNe) are produced per stellar mass formed, assuming a Salpeter $1-100 \msun$ initial mass function, and each of them yields $y_d\simgt 0.1 \msun$ of dust \citep{Ferrara22}, then we would expect a dust mass $M_d \simgt 3\times 10^6 \msun$ to be present in the system\footnote{Note that this argument neglects dust growth in the interstellar medium. Although likely negligible \citep{Ferrara16}, the above value then represents a strong lower limit on $y_d$.}. This corresponds to a dust-to-stellar mass ratio of $\simgt 1/529$, about a factor $\approx 2 \times$ larger than in the Milky Way (MW).  

By further assuming that $\approx 10$\% of the baryonic fraction, $f_b = \Omega_b/\Omega_m$, of their halos ($M\approx 10^{11.3} \msun$, \citealt{Ferrara22a}) cools in the disk, we find a gas-to-stellar ratio of $M_g/M_*=2.1$. This corresponds to a gas fraction $f_g = M_g/(M_g+M_*)= 0.68$, and a dust-to-gas ratio $D \approx 0.001$, i.e. 6 times lower than in the MW. As \citet{Remy14} have shown that $D \propto D_{\rm MW} (Z/Z_{\odot})$ in the range $0.1< Z/Z_{\odot}<0.5$ expected for JWST galaxy candidates \citep{Tacchella22, Curti22}, $D=0.001$ is a reasonable choice. Although uncertain, in the following we will use these values for $f_g$ and $D$ as an educated guess for the super-early galaxy population.

Assuming a spherical system with an effective radius $r_e$, the dust optical depth at 1500\AA\, is  
\begin{equation}
\tau_{1500} = \frac{\sigma_{1500}}{4\pi r_e^2 m_p D_{\odot}} M_d,    
\label{eq:tau1500}
\end{equation}
where $\sigma_{1500}=1.3\times 10^{-21}(D/D_\odot)$ is the dust extinction cross section, $m_p$ is the proton mass, and $D_{\odot}=1/162$ is the MW dust-to-gas ratio. Eq. \ref{eq:tau1500} numerically yields 
\begin{equation}
\tau_{1500} \simgt 6.25 \left(\frac{M_d}{3\times 10^6 \msun}\right) \left(\frac{{\rm kpc}}{r_e}\right)^{2}.    
\label{eq:tau1500_num}
\end{equation}
As at these high redshifts $r_e \approx 0.5$ kpc for a galaxy with stellar mass $M_* \approx 10^9 \msun$ \citep{Pallottini22, Adams22}, the above calculations show that we should expect these massive galaxies to be heavily obscured, i.e. $\tau_{1500} \approx 25$. How do we overcome this apparently conflicting result? There are two possibilities that we will explore in sequence in the next two Sections. 

The first solution is that the dust produced by the SNe on very short time scales (typically, $5-10$ Myr after the onset of the star formation activity) is ejected into the intergalactic medium at a rate exceeding its production rate. Evacuation can take place, for example, via the strong radiation pressure exerted by the observed UV-emitting stars. Of course, it is necessary to clarify if and under what conditions such process can effectively work.

An alternative explanation involves spatial segregation between the UV-emitting regions and the dust. This UV-dust continuum displacement is relatively common in lower redshift ($z \approx 6-7$) galaxies, as shown by a number of observations and models \citep[][]{Carniani18, Bowler18, Cochrane19, Zanella21, Inami22}. In this case the freshly produced dust remains in the system, but it  only marginally attenuates the UV light emitted by stars. 

Remarkably, these two scenarios make distinct predictions concerning the dust continuum luminosity of the JWST-detected galaxies typically measured at 158 $\mu$m restframe wavelength.  While ejection entails a very low $F_{158}$ flux, due to the lack of significant amounts of dust in the galaxy, the spatial segregation scenario predicts that these systems might be noticeable FIR emitters. In the following we quantify these statements and provide further arguments.

%
% FIGURE 1
%

\section{Dust ejection scenario}\label{sec:eje}
We explore the possibility that dust is ejected via a galactic outflow driven by UV radiation pressure on grains\footnote{We assume perfect dynamical coupling between dust and gas.}, assuming a disk galaxy. The radiation pressure is given by 
\begin{equation}
p_r=(1-e^{-\tau_{1500}})\frac{\mathcal{F}_{\rm bol}}{c},
\label{eq:pr}
\end{equation}
where $c$ is the speed of light.
Following \citet[][]{Ferrara19}, we write the bolometric flux\footnote{This expression is valid for a stellar population with a $1-100 M_\odot$ Salpeter IMF, $Z = 0.5 Z_\odot$ at age of 10 Myr.} (erg cm$^{-2}$ s$^{-1}$) as $\mathcal{F}_{\rm bol}= 5\ \Sigma_{\rm SFR}[M_\odot {\rm yr}^{-1} {\rm kpc}^{-2} ]$\footnote{This relation implies a number of ionizing photons per unit SFR, $N_i=10^{53.4}$, obtained from the population synthesis code STARBURST99 by assuming an average age of $100\, \rm Myr$, and a standard Salpeter IMF. In the metallicity range discussed in Sec. \ref{sec:prel}, $N_i$ can vary within a factor $<3$.}, 
where $\Sigma_{\rm SFR}$ is the disk star formation rate (SFR) per unit area. 
The gravitational pressure of the gas at the disk mid-plane is instead
\begin{equation}
p_g=\frac{\pi}{2}G\frac{\Sigma_g^2}{f_g},
\label{eq:pg}
\end{equation}
where $G$ is the gravitational constant, and $\Sigma_g$ is the gas surface density.

To drive an outflow, the radiation pressure must overcome the gravity pressure. By imposing that the Eddington ratio, $\lambda_E = p_r/p_g > 1$, we then derive the necessary condition for an outflow to develop as a function of $\Sigma_{\rm SFR}$, and burstiness parameter, $\ks$, implicitly defined \citep[][]{Ferrara19} by the following expression:
\be
\Sigma_{\rm SFR}= 10^{-12}\, \ks {\Sigma_g}^{1.4}.
\label{KS}
\ee
Physically, $\ks$ quantifies deviations from the local \citep[][]{Heiderman10} Kennicutt-Schmidt relation. 

\begin{figure}
\centering\includegraphics[width = 0.46 \textwidth]{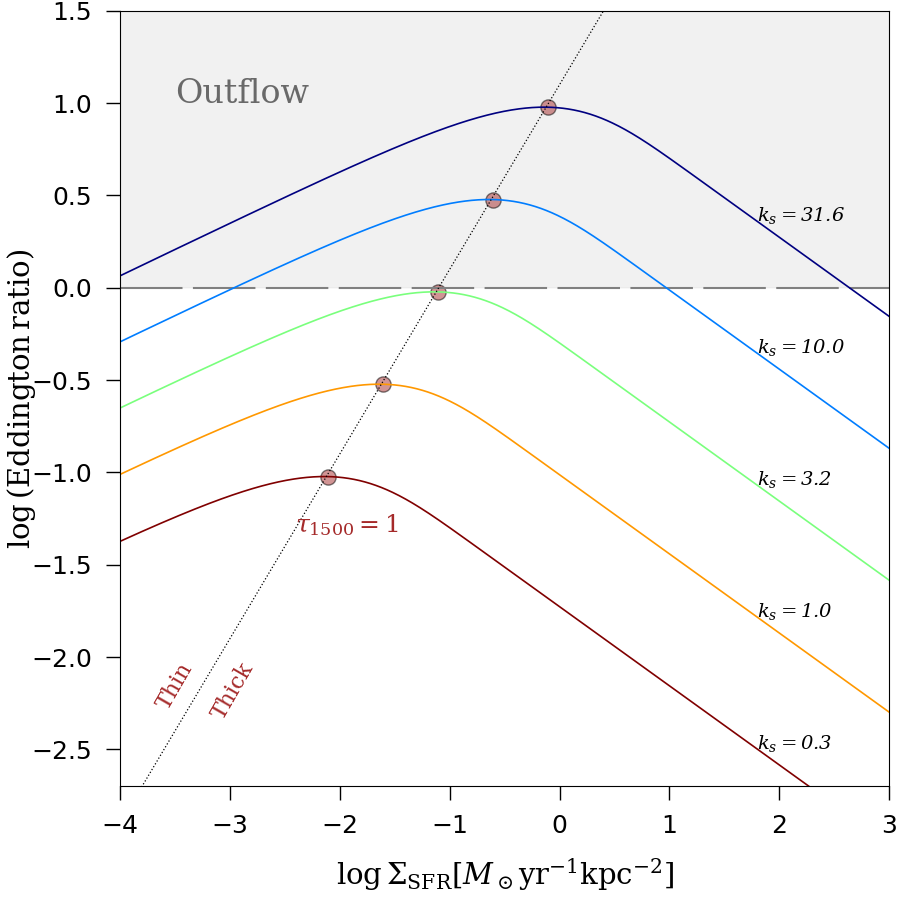}
\caption{Eddington ratio, $\lambda_E \equiv p_r/p_g$, as a function of $\Sigma_{\rm SFR}$. Curves corresponds to different $\ks$ values, as indicated. Outflows develop in the grey region where $\lambda_E > 1$. Red dots mark the locus $\tau_{1500}=1$, which maximises $\lambda_E$. The grey dotted line separates optically thin/thick regimes.}
\label{Fig:02}
\end{figure}

The behavior of $\lambda_E$ with $\Sigma_{\rm SFR}$ for various $\ks$ values is shown in Fig. \ref{Fig:02}. The grey area identifies the region of the parameter space in which an outflow is expected ($\lambda_E > 1$).  
Independently of $\Sigma_{\rm SFR}$, there is a limiting value, $\ks^\star \simeq 3.3$, below which outflows cannot be launched. This is obtained by imposing $\lambda=1$, and solving for $\Sigma_{\rm SFR}$ and $k_s$. A solution always exists for $k_s > 3.3$; below that threshold, no solution can be found.

By combining eq. \ref{eq:pr} and \ref{eq:pg} we find that outflows can occur, for $\ks > \ks^\star$ and $D=0.001$, in the $\Sigma_{\rm SFR}$ range
\begin{equation}
    0.32\, \ks^{-5/2}  \le \Sigma_{\rm SFR} \le 4.3\times 10^{-3} \ks^{10/3}.
\end{equation}

The trend of the Eddington ratio can be  understood as follows. In the optically thin regime, $p_r \sim \tau_{1500} \Sigma_{\rm SFR} \sim \Sigma_g  \Sigma_{\rm SFR}$; hence $\lambda_E \sim \Sigma_{\rm SFR}^{2/7}$, having used the KS relation (eq. \ref{KS}). In the optically thick regime ($\tau_{1500}\gg 1$), it is instead $\lambda_E \sim \Sigma_{\rm SFR}^{-3/7}$. Thus, $\lambda_E$ has a maximum located at $\tau_{1500} \approx 1$, depicted by the red dots in Fig. \ref{Fig:02}.

%
% FIGURE 2
%
%

%
% FIGURE 3
%
\begin{figure}
\centering\includegraphics[width = 0.46 \textwidth]{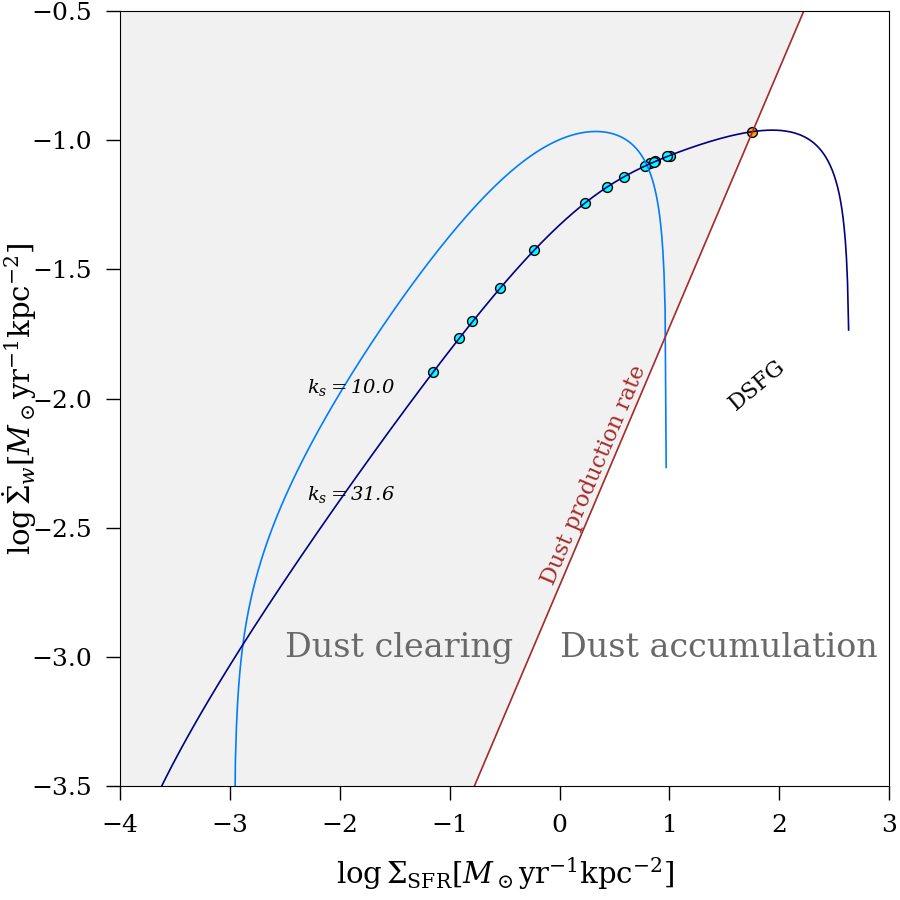}
\caption{Dust outflow rate, $\dot \Sigma_w$, as a function of $\Sigma_{\rm SFR}$ for $z=11.5$, assuming $y_d = 0.1\ \msun$. Shown are two curves corresponding to $\ks=(10, 31.6)$ for which radiatively-driven outflows can occur in the displayed $\Sigma_{\rm SFR}$ range. When the curves are located in the \quotes{dust clearing} region, the dust ejection rate is faster than the production rate (red curve) and the galaxy is cleared; in the opposite case (white region) dust accumulates as a result of a too slow ejection rate. Shown for comparison, arbitrarily assuming $\ks=31.6$ appropriate for the starburst nature of these galaxies, are the almost dust-free $z>10$ candidates reported by \citet{Furtak22} (cyan points), and the obscured ($A_{\rm V} = 2.5$) galaxy by \citet{Rodighiero22} at $z=12.1$ (orange), for which an effective radius of $0.5$ kpc has been assumed  to compute $\Sigma_{\rm SFR}$. }
\label{Fig:03}
\end{figure}

The condition $\lambda_E>1$ is not sufficient to guarantee that dust is fully evacuated. To complete the dust clearing process, the dust ejection rate $\dot{\Sigma}_w$ must exceed the dust production rate $\dot{\Sigma}_{\pi}$.
In the super-Eddington regime ($p_r>p_g$), the radiation pressure accelerates the gas to a velocity $v$ (perpendicular to the disk) given by momentum conservation:
\begin{equation}
    \Sigma_g\frac{dv}{dt}=p_r-p_g.
\end{equation}
The gas velocity at the disk scale height $H$ is then
\begin{equation}\label{vel}
    v_{\infty}=\sqrt{\frac{2H}{\Sigma_g}(p_r-p_g)},
\end{equation}
with a gas outflow rate, $\dot{\Sigma}_g$, given by
\begin{equation}
    \dot{\Sigma}_g v_{\infty}=(p_r-p_g),
\end{equation}
while the gas scale height can be expressed as
\begin{equation}
    H=\frac{\sigma^2}{\pi G\Sigma}
\end{equation}
where $\sigma$ is the turbulent r.m.s. velocity dispersion, which is related to the SFR by $\sigma=\Sigma_{\rm SFR}/ \zeta^2$, with  $\zeta=0.041(1+z)^{3/2} f_g$, representing the rotational frequency of a galaxy hosted by a dark matter halo of mass $M_h$ (Ferrara, in prep.).

Finally, $\Sigma=\Sigma_g/f_g$ is the total (gas+stars) mass surface density.
By combining the last two expressions we obtain the dust ejection rate\footnote{As $\dot{\Sigma}_w \propto D$, the normalization of the curves in Fig. \ref{Fig:03} may vary by a factor $\approx 3$ due to uncertainties in metallicity measurements.}
\begin{equation}\label{ejrate}
    \dot{\Sigma}_w=\frac{1}{2}v_{\infty}\frac{\Sigma_g}{H}D.
\end{equation}
%\AF{Zip, ma rileggi o no? Dove e' definito H?} 
The dust production rate can be derived from the SFR surface density as:
\begin{equation}
    \dot{\Sigma}_{\pi}= y_d \nu_{\rm SN}\Sigma_{\rm SFR}.
\end{equation}

{In principle, two mechanisms can destroy dust. These are: (a) destruction by the reverse shock processing the SN ejecta; (b) destruction of pre-existing ISM dust by the SN shock. We implicitly account for (a) by adopting a dust net yield/SN of $y_d = 0.1 \msun$, corresponding to the destruction of $\sim 90$\% of the dust freshly produced by the SN \citep{Matsuura19, Lesniewska19, Slavin20}. For simplicity, however, we do not include the effects of (b). Neglecting this effect is a conservative assumption, as adding ISM dust destruction would result in a lower $\Sigma_{\rm SFR}$ threshold for the dust clearing regime. Also, dust destruction due to thermal sputtering is inefficient at $T \simlt 10^5$ K. As it is likely that only a small fraction of the galaxy volume is filled with hotter gas, this effect is expected to be small, as indeed recently found by \citet{Nath22}.}

Fig. \ref{Fig:03} shows the outflow rate as a function of $\Sigma_{\rm SFR}$.
The blue (purple) curve represents the case $\ks=10$ ($\ks=31.6$).
The dust production rate (red curve) separates the two regimes of dust clearing and dust accumulation. 
In the clearing region (grey shaded area) the dust ejection rate overcomes its production rate, and dust is efficiently evacuated into the circumgalactic medium. Outside that region, dust ejection becomes inefficient, and dust accumulates in the galaxy. 

%Above the critical value $\ks^{\star}$, it is possible to identify a threshold value $\Sigma_{\rm SFR}^{\star}$ below which the condition $\dot{\Sigma}_w>\dot{\Sigma}_{\pi}$ is always satisfied.
%We define this threshold value as
%\begin{equation}
    %\Sigma_{\rm SFR}^{ \star}= ..... TBD.
%\end{equation}

%As shown in fig. \ref{Fig:02}, only a part of the super early galaxies are dust transparent. The two cases (a) $\ks < \ks^\star$ and (b) $\Sigma_{SFR} >\Sigma_{SFR}^\star$ define which galaxies are unable to complete the dust ejection. In those regimes, we expect early galaxies to have large ($>1$) dust attenuation, $A_{\rm V}$. 
%\AF{Next sentence does not connect with the dust accumulation point before!}

{The majority ($\sim 60\%$)} of JWST super-early sources show a very blue ($\beta <-2$) UV slope, which might result from dust ejection. As we have seen, ejection requires specific conditions highlighted in Fig. \ref{Fig:02} which might not be met by all high-$z$ galaxies. \citet{Rodighiero22} studied a sample of very red galaxies ($A_{\rm V}\sim 5$) which are detected in the F444 band, but missed in F200. Remarkably, they find two strongly dust-attenuated galaxies at $z>10$, thus confirming that dust accumulation is active also at these very early epochs. We warn that the identification of the observed break as a standard Lyman-break can be misintepreted as dusty star-forming lower redshift ($z<7$) interlopers\footnote{\citet{Naidu22b} and \citet{Zavala22} presented similar cases in which an extremely high redshift ($z\sim 12-17$) candidate has a second interpretation with a lower redshift ($z=5$) solution.}.

\section{Dust spatial segregation scenario}\label{sec:bench}
We now discuss an alternative scenario to explain the low $A_{\rm V}$ observed in $z \simgt 10$ galaxies, which involves the \quotes{spatial segregation} of stellar (optical-to-UV) and dust continuum (infrared, IR) emitting regions 
\citep{Behrens18,Sommovigo20,sommovigo22b,Ferrara22,Pallottini22,Dayal22}. In this scenario, the UV radiation mostly comes from the transparent diffuse interstellar medium (ISM), hosting either little or cold dust. The dust-obscured SFR is instead located in giant molecular clouds (GMCs), strongly emitting at IR wavelengths.

% observable: Im
\cite{Ferrara22} introduced a quantitative measure of the ISM morphology of the ISM, the molecular index $I_{m}$. This is defined as:
\begin{equation}\label{eq:Im}
    I_{m} = \frac{(F_{\rm IR}/F_{\rm UV})}{(\beta-\beta_{\rm int})},
\end{equation}
where $F_{\rm IR}$ ($F_{\rm UV}$) is the monochromatic flux at rest-frame $158 \mathrm{\mu m}$ ($1500 \angstrom$).
For a single zone, optically-thin ISM, \cite{Ferrara22} obtains an analytical expression for $I_{m}$ which shows a maximum\footnote{The presence of a maximum value for $I_{m}$ in an  optically-thin medium has a straightforward physical explanation. $F_{\rm IR}$, and thus $I_{m}$, can be increased by raising either the dust temperature, $T_{\rm d}$, or mass, $M_{\rm d}$. However, increasing $T_{\rm d}$ requires a larger $A_{\rm V}$, which is excluded in an optically-thin medium. Raising $M_{\rm d}$ while keeping $A_{\rm V}$ low is possible, but it implies pushing $T_{\rm d}$ progressively closer to the CMB temperature. This prevents $F_{\rm IR}$, and thus $I_{m}$, to increase indefinitely. Hence, $I_{m}> I_{m}^*$ values can only be attained in a multi-phase ISM, where UV and IR emission are essentially decoupled. This is the case of spatially-segregated sources.} at $I_{m}^* \simeq 1120$.  

\begin{figure}
\centering\includegraphics[width = 0.46 \textwidth]{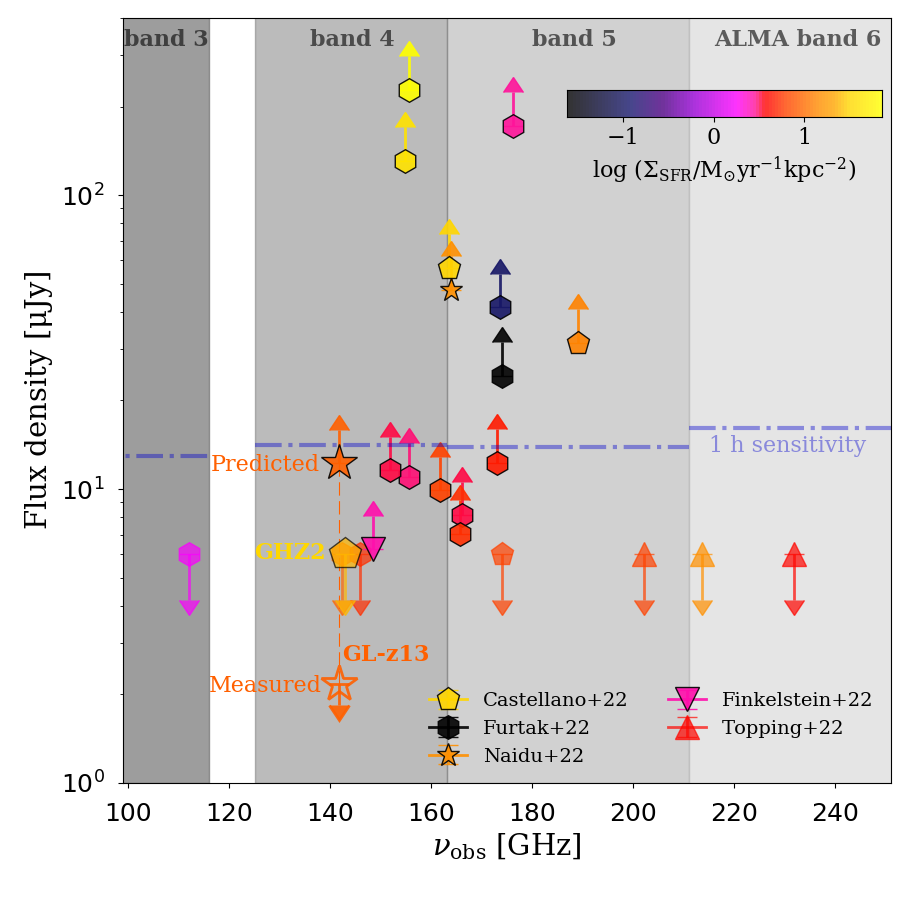}
\caption{Predicted FIR continuum flux at rest-frame $158\ \mathrm{\mu m}$ as a function of the corresponding observed frequency $\nu_{\rm obs}$ for the $z>9.5$ galaxies detected so far with JWST \citep{Atek22,Naidu22a,Finkelstein22, Robertson22}. The galaxies are colour-coded according to their $\Sigma_{\rm SFR}$ as in Fig. \ref{Fig:01}. The lower limits for $F_{\rm IR}$ shown here are obtained as described in Sec. \ref{sec:bench}, under the assumption of all the sources being spatially-segregated. For sources with $\beta<\beta_{\rm int}$, an upper limit for $F_{\rm IR}$ is fixed at the minimum value of the sample ($\sim 6\ \mathrm{\mu Jy}$). The ALMA bands in which $F_{\rm IR}$ would be measured are also shown, alongside the sensitivity reached with 1 hour of observation. For GHZ2/GL-z13 we show the upper limit for the $158\ \mathrm{\mu m}$ continuum flux inferred by rescaling the $3\sigma$ upper limit measured by \cite{Bakx22} at rest-frame $88\ \mathrm{\mu m}$ (assuming $T_{\rm d}=50\ \mathrm{K}$ and MW dust).}
\label{Fig:04}
\end{figure}

% spiego Fig. 4
Under the assumption that the nearly dust-unattenuated super-early galaxies observed with JWST are spatially segregated systems ($I_{m}>I_{m}^*$), we can make a testable prediction on their $F_{\rm IR}$. Inverting eq. \ref{eq:Im}, from the observed $F_{\rm UV}$ we infer a lower limit to their $F_{\rm IR}$. This is shown in Fig. \ref{Fig:04}. Thanks to the very high-$z$ of the considered sources, their rest-frame $158\mathrm{\mu m}$ continuum emission is redshifted at $> 1\ \mathrm{mm}$ wavelengths, traced by the sensitive ($\approx 13\ \mathrm{\mu Jy}$ in 1 hr of observation) ALMA bands 3, 4, 5 and 6. We predict $F_{\rm IR}> 6\ \mathrm{\mu Jy}$ for all of the considered sources with $\beta>-2.616$; thus, few hours of observations with ALMA would suffice to test the validity of the spatially segregated scenario. This is a noticeable example of the leap forward in terms of high-$z$ galaxies characterisation that will soon be possible thanks to ALMA and JWST synergy. {For the sources where the inferred UV slope is bluer than the assumed intrinsic one, the previously described method to predict $F_{\rm IR}$ cannot be applied. In these cases, however, we do expect an extremely low dust content, and thus negligible FIR emission, consistent with the dust ejection scenario.}

\citet[][]{Bakx22} and \citet[][]{Popping22} performed deep spectroscopic and continuum ALMA observation of the $z\simeq 12$ galaxy GHZ2/GL-z13 \citep{Castellano22,Naidu22a}, searching for \OIII line emission at rest-frame $88 \mathrm{\mu m}$ (see also \citealt{Fujimoto22, Kaasinen22, Yoon22}). They fail to detect the $88 \mathrm{\mu m}$ dust continuum emission from this source, obtaining a $3\sigma$ upper limit of $<13.8\ \mathrm{\mu Jy}$. 

Assuming a dust temperature of $50\ \mathrm{K}$, we can rescale this value to obtain an upper limit on the continuum flux at $158\ \mathrm{\mu m}$, $F_{\rm IR}<2\ \mathrm{\mu Jy}$ (open orange star in Fig. \ref{Fig:04}). 
{Such $F_{\rm IR}$ value is $\times 6$ smaller than the lower limit ($F_{\rm IR} \geq 12\ \mathrm{\mu Jy}$, orange star in Fig. \ref{Fig:04}) we predict for this source when considering $\beta=-2.4$ \citep{Bakx22}. We note that if we consider the most extreme $\beta=-3$ inferred by \cite{Castellano22,Santini22} for this same source, we can only provide a qualitative upper limit on $F_{\rm IR}<6\ \mathrm{\mu Jy}$, which is consistent with the ALMA non-detection. We interpret this result as an indication that this galaxy is dust-cleared by an outflow.}

As a final remark, the spatially segregated, multi-phase ISM scenario is also disfavoured by the very compact size of GHZ2/GL-z13 ($\approx 100\ \mathrm{pc}$, \citealt{Bakx22}). Further ALMA observations of the other super-early galaxies will make our prediction more robust.\\

%
% FIGURE 4
%

\section{Summary and discussion}

To explain the puzzling evidence reported by JWST Early Release Science surveys for \quotes{blue monsters}, i.e. super-early ($z>10$), massive galaxies showing very blue spectra ($\beta \simeq -2.6$) and dust attenuation ($A_{\rm V} \simlt 0.02$), we have proposed two distinct scenarios. 

\textit{Dust ejection.} In the first scenario, $A_{\rm V}$ is small as a major fraction of the dust content is ejected from the galaxy by UV radiation pressure produced by newly born, massive stars. This process requires that the galaxy is in a starburst phase ($\kappa_s > \kappa_s^{\star} = 3.3$), and that the dust ejection rate exceeds its production one by SN. Such condition is shown in Fig. \ref{Fig:03}. Galaxies with very large SFR surface density can instead accumulate dust and become significantly obscured, perhaps explaining the recent detection of $A_{\rm V} \approx 5$ galaxies in the same JWST data \citep[][]{Rodighiero22}.

\textit{Dust spatial segregation.} The alternative scenario involves a spatial segregation of dust (mostly located in opaque GMCs) with respect to transparent UV-emitting regions. This scenario can explain the blue spectra colors, and at the same time it makes it possible to predict the FIR continuum emission from these sources using eq. \ref{eq:Im}, once the slope $\beta$ is known. 

Although both scenarios provide a physical explanation for the blue colors of the observed galaxies, they make distinct prediction concerning the FIR dust continuum flux observed by e.g. ALMA. 
If most of the dust is ejected in the circumgalactic medium, both its density and temperature drop dramatically, making its detection virtually impossible. On the other hand, the segregation scenario should result in (rest-frame) 158 $\mu$m fluxes $F_{\rm IR} > 6\, \mu$Jy {for most sources}, which are detectable by ALMA in bands 3, 4, 5, and 6 in a few hours. 

Recently, \citet[][]{Bakx22, Popping22} reported results from continuum ALMA observations towards GHZ2, one of the brightest and most robust candidates at $z>10$, identified in the GLASS-JWST Early Release Science Program \citep[][]{Treu22}. They obtained an upper limit for the flux at (restframe) 88 $\mu$m flux of $ 13\ \mu$Jy. This non-detection favours the ejection scenario, and it implies a stringent upper limit on the dust mass $< 6.5\times 10^6 \msun$, assuming a dust temperature of 50 K.

If the dust is indeed efficiently ejected during the starburst phase of (some of) the super-early galaxies, such outflow might also carry a significant fraction of the gas to which grains are likely to be tightly dynamically coupled. These galaxies might then also retain a very limited amount of gas. This fact has two important consequences: (a) the escape fraction of ionizing photons could be very high; (b) the production of nebular lines can be highly suppressed. It is then tantalising to interpret the non-detection of [OIII] 88$\mu$ line in GHZ2/GL-z13 \citet[][]{Bakx22} in this physical framework (see \citealt{Kohandel22} for a detailed discussion). Noticeably, \citealt{Kaasinen22, Yoon22} also attempted at observing the [OIII] 88$\mu$ line for two JWST $z>10$ galaxy candidates (HD1, GHZ1) without any success. 
% makes a graphical mess if it is after the \acknowledgments
\section*{Data Availability}
%{\bf Data Availability}\\
Data available on request.

\acknowledgments
We thank S. Carniani, M. Castellano, S. Gallerani, L. Pentericci, G. Rodighiero, P. Rosati, P. Santini, E. Vanzella for useful discussions. AF, LS, MK acknowledges support from the ERC Advanced Grant INTERSTELLAR H2020/740120. Generous support from the Carl Friedrich von Siemens-Forschungspreis der Alexander von Humboldt-Stiftung Research Award is kindly acknowledged (AF). 
Plots in this paper produced with the \textsc{matplotlib} \citep{Hunter07} package for \textsc{PYTHON}.    

\bibliographystyle{aasjournal}
\bibliography{paper}



\end{document}